# Millimeter Wave Beamforming Based on WiFi Fingerprinting in Indoor Environment

[1,2]Ehab Mahmoud Mohamed, [1]Kei Sakaguchi, and [1]Seiichi Sampei
[1]Graduate School of Engineering, Osaka University, [2]Electrical Engineering Dept., Aswan University.
Email: ehab@wireless.comm.eng.osaka.ac-u.ac.jp, {sakaguchi, sampei}@comm.eng.osaka-u.ac.jp

*Abstract*—**Millimeter Wave (mm-w), especially the 60 GHz band, has been receiving much attention as a key enabler for the 5G cellular networks. Beamforming (BF) is tremendously used with mm-w transmissions to enhance the link quality and overcome the channel impairments. The current mm-w BF mechanism, proposed by the IEEE 802.11ad standard, is mainly based on exhaustive searching the best transmit (TX) and receive (RX) antenna beams. This BF mechanism requires a very high setup time, which makes it difficult to coordinate a multiple number of mm-w Access Points (APs) in mobile channel conditions as a 5G requirement. In this paper, we propose a mm-w BF mechanism, which enables a mm-w AP to estimate the best beam to communicate with a User Equipment (UE) using statistical learning. In this scheme, the fingerprints of the UE WiFi signal and mm-w best beam identification (ID) are collected in an offline phase on a grid of arbitrary learning points (LPs) in target environments. Therefore, by just comparing the current UE WiFi signal with the pre-stored UE WiFi fingerprints, the mm-w AP can immediately estimate the best beam to communicate with the UE at its current position. The proposed mm-w BF can estimate the best beam, using a very small setup time, with a comparable performance to the exhaustive search BF.**

## I. INTRODUCTION

Millimeter-wave (mm-w) band communications, particularly the 60 GHz band, have received a considerable attention as a key enabler for the 5G cellular networks [1] - [3]. The most attractive feature of the mm-w communication is its ability to attain multi-Gbps rate, which can increase the cellular network capacity to sustain the expected huge increase in mobile data traffic. However, mm-w communication suffers from high propagation loss due to its high frequency band. In order to compensate the tremendous propagation loss and reduce the shadowing effect, high-gain directional antenna array is favored to improve the system efficiency and transmission range thanks to the millimeter range of the antenna size and spacing.

Beamforming (BF) determines the best beam direction formed by multiple antenna elements to maximize the transmission rate. Mm-w BF schemes based on estimating the entire channel state information (CSI) suffer from high calculation load and large overhead [4]. Instead, various Medium Access Control (MAC) based BF protocols have been proposed for mm-w transmissions, in which, switched antenna array with a structured codebook is used. The main MAC based BF protocols proposed for mm-w communications are the beam codebook [4], the iterative search [5] and the multiple sector ID capture (MIDC) [6]. The BF protocol proposed by the IEEE 802.11ad standard is mainly based on the MIDC protocol, which is an exhaustive search BF protocol [7].

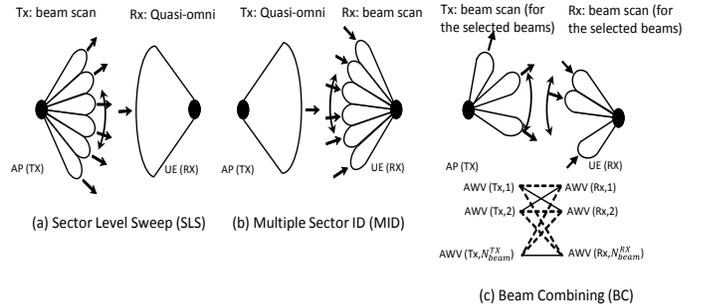

Fig. 1. The IEEE 802.11ad (MIDC) BF protocol.

The MIDC BF protocol mainly consists of the Sector Level Sweep (SLS) phase, the Multiple Sector Identifier (MID) subphase and the Beam Combining (BC) subphase [6]. The MIDC BF protocol for downlink transmission (AP-UE link) is shown in Fig. 1. In the SLS phase, the mm-w TX (AP) beam is scanned, by changing antenna settings (Antenna Weight Vectors, AWVs), where the RX (UE) beam on the other link end is kept in quasi-omni mode, as illustrated in Fig. 1 (a). On the other hand, in the MID subphase, the mm-w RX (UE) beam is scanned while the TX (AP) beam is kept in quasi-omni, as illustrated in Fig. 1 (b). At the end of the SLS and MID, TX and RX beam candidate tables are constructed. These tables contain the TX and RX antenna sector IDs (AWVs) corresponding to higher quality TX and RX links [6]. Using these TX and RX candidate beams, a beam combining subphase (BC) is performed in a round robin trail, Fig. 1 (c). The BC subphase is used to overcome the problem of imperfect quasi-omni antenna pattern, which cannot be avoided using the beam codebook or the iterative search BF protocols [6]. As a result of the BC subphase, a table of multiple antenna setting pairs (ASPs) corresponding to higher link qualities is reserved. This allows for fast beam switching when link blockage occurs between the TX and the RX. According to [6], using 32 different antenna sector IDs in the SLS phase and the MID subphase and 7 beams for the BC subphase, the typical BF setup time consumed by a mm-w AP using the MIDC BF protocol is about 1.8 msec. The SLS phase consumes more than 70 % of the total BF time. This is because control packets are used in the SLS phase, where one preamble must be transmitted per one antenna setting [6] [7].

For 5G mobile applications, a large number of mm-w APs should communicate concurrently with their associated UEs using coordinated transmissions. Coordinated mm-w concurrent transmissions should be used to fully cover a typical indoor environment while maximizing total system capacity, supporting more users, reducing user outage rate and assuring fairness among users especially for densely populated networks. Seamless handover should be also coordinated among the short range mm-w APs especially for large enterprise scenarios. Coordinated mm-w transmissions make BF even more difficult, because at every decision of coordinated transmissions, the best beams for all AP–UE links should be known beforehand. For example, the mutual interferences (available data rates), provided by all alternative AP-UE combinations, should be known before making a decision for joint user scheduling while maximizing the total system rate [2]. These available data rates are unknown until the best transmission and reception beams for all AP-UE links are finalized [2]. The use of conventional BF (SLS, MID and BC) will result in very high system setup time because all APs will exhaustively search the best beams with their associated UEs at every time slot of coordinated concurrent transmissions. The use of conventional BF even becomes unrealistic if we try to coordinate a high number of mm-w APs with a high number of antenna sectors in mobile channel conditions, which is a typical assumption for the future 5G networks. Using 32 TX and RX antenna sectors and 7-beam for BC, 10 mm-w APs will consume more than 18 msec for finalizing the best beams. This BF setup time will be increased when using a higher number of antenna sectors, and it should be consumed at every time slot of mm-w coordinated concurrent transmissions.

In this paper, we propose a novel BF protocol that greatly reduces the BF setup time with a comparable performance to the exhaustive search MIDC protocol. The proposed BF is based on statistical learning, in which we make use of the wide coverage WiFi (5 GHz) fingerprinting to localize mm-w (60 GHz) best sector IDs in indoor environment. Therefore, by just comparing current UE WiFi signal with the pre-stored UE WiFi fingerprints, a mm-w AP can immediately estimate the best beam to communicate with the UE at its current position using the pre-localized best sector IDs. In this paper, we assume the use of dual band (5 GHz / 60 GHz) UEs, which is a typical assumption for future 5G networks. The UE WiFi signal can be received by the mm-w APs themselves using dual band (5GHz / 60 GHz) mm-w APs, or using separate 5 GHz (WiFi) APs. By focusing on the main concept without losing the generality, we use Received-Signal-Strength (RSS) of the UE WiFi signal as a simple fingerprint of the WiFi signal. Other WiFi fingerprinting techniques such as Time Difference of Arrival (TDOA), Direction of Arrival (DOA) and Channel State Information (CSI) can be easily used by extending the proposed BF [8].

Simulation analysis confirms the high efficiency of the proposed BF protocol in estimating the best beam compared to the exhaustive search BF. In addition, a great reduction in the BF setup time is obtained using the proposed BF.

The rest of this paper is organized as follows; Section II provides the proposed system model. The proposed BF mechanism is presented in Sect. III. The performance of the proposed BF protocol is analyzed in Sect. IV via simulation analysis. Section V concludes this paper.

## II. THE PROPOSED SYSTEM MODEL

Figure 2 shows the details of the proposed system model. For the purpose of generalization, we assume separate deployments for the 5 GHz (WiFi) APs although dual band (5 / 60 GHz) mm-w APs can be used. The 60 GHz mm-w APs and the 5 GHz WiFi APs are connected to a local controller. The local controller can be implemented as an AP controller for mm-w APs and WiFi APs. In Fig. 2, the WiFi APs and mm-w APs are connected to the controller via optical fiber links or gigabit Ethernet. This system will be installed in a target environment to cover it with the high- capacity mm-w APs. The WiFi and mm-w best sector ID fingerprints are collected and stored in the controller in an offline phase.

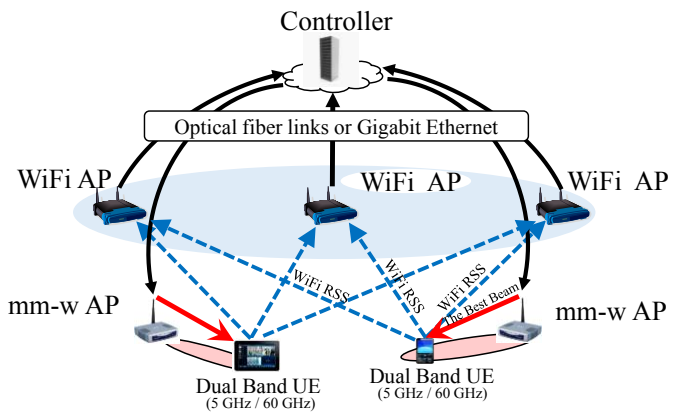

Fig. 2. The proposed system architecture.

In addition, the controller performs grouping and clustering on the collected WiFi fingerprints to find out the best WiFi fingerprint exemplars that can effectively localize each mm-w best sector ID.

In the online BF phase, after comparing current UE WiFi readings with the pre-stored WiFi exemplars, group of best sector IDs (best beams) are estimated for a mm-w AP-UE link. A beam combining (BC) subphase is conducted using the estimated beams to find out the best beam with the highest link quality. Beside overcoming the problem of imperfect quasi-omni antenna pattern, the BC subphase is mainly used by the proposed BF protocol to alleviate the real-time beam blocking, e.g., human shadowing, which may not occur in the offline phase. Also, it is used to overcome the inaccuracy in WiFi RSS measurements.

In addition to the BF functionality, the controller can be used as a coordinator to coordinate the transmissions among mm-w APs and WiFi APs such as performing association, re-association, joint user scheduling... etc. To facilitate the coordination functionality, the concept of control plane / user plane (C/U) splitting can be used. In which, control signals are sent to the UEs using the wide coverage WiFi APs and a combination of mm-w APs and WiFi APs are used to deliver the data. Also, switching ON / OFF functionality can be added to the controller to reduce the total energy consumption. Thus, the controller can switch ON / OFF the mm-w APs based on the usage. In addition, it can switch ON / OFF the UE mm-w interface, using the wide coverage WiFi signaling, based on the usage of mm-w link. Likewise, the controller works as a Gateway to connect the proposed mm-w / WiFi system with the global network. The proposed WiFi / mm-w coordination system can be considered as an enabler for 5G cellular networks as well as for future wide coverage multi-gigabit WLAN.

## III. THE PROPOSED BEAMFORMING MECHANISM

In this section, we give the details of the proposed mm-w BF mechanism. The proposed mechanism is used to effectively estimate the best beam for a mm-w AP-UE link using a statistical learning approach. The idea behind the proposed BF mechanism is to eliminate the SLS phase, which consumes the highest setup time, from the real-time BF. SLS elimination can be effectively done by performing it in an offline phase and positioning the mm-w best sector IDs using WiFi fingerprints. Consequently, the real-time BF can be easily done by just comparing the current UE WiFi readings with the pre-stored WiFi fingerprints. As a result, a very small setup time BF protocol is obtained with a comparable performance to the exhaustive search BF. Figure 3 shows the general framework of the proposed BF for one mm-w AP-UE link, which will be explained in more details through the subsequent sub-sections.

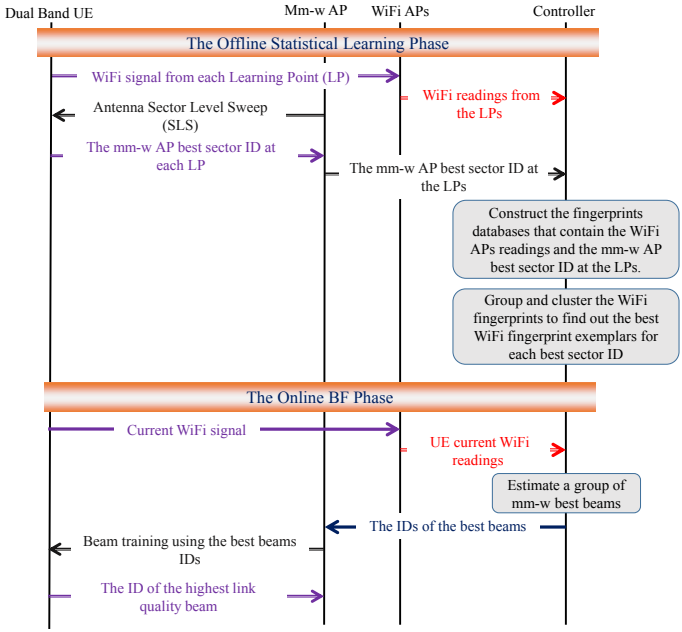

Fig. 3. The proposed beamforming mechanism.

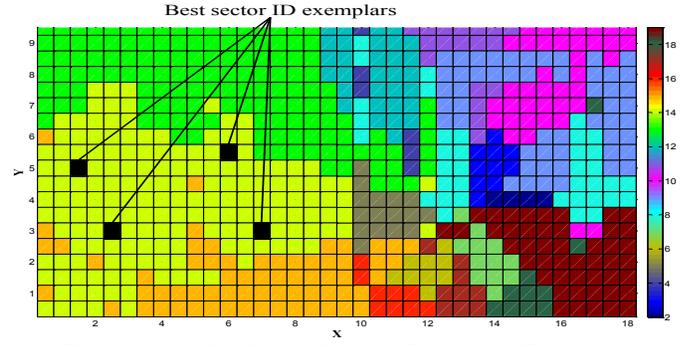

Fig. 4. An example of a typical mm-w AP best sector ID radio map.

## A. The offline Statistical Learning Phase

### 1) Collecting Fingerprints Databases (DBs)

The first step in the offline statistical learning phase is to construct the micro-wave and mm-wave radio maps for the target environment. For the sake of simplicity, we will use the Received Signal Strength (RSS) as a fingerprint of the WiFi signal; however, other WiFi fingerprinting methodologies can be easily used without any modifications to the general concept of proposed BF protocol. Constructing the radio maps can be effectively done by collecting the average WiFi RSS readings and mm-w APs best sectors IDs at arbitrary Learning Points (LPs) in the target environment. Therefore, two databases are constructed as the fingerprinting radio maps, named the WiFi RSS DB $\Psi$, and the best sector ID DB $\Phi$, which can be defined as:

$$\Psi = \begin{pmatrix} \psi_{11} & \cdots & \psi_{L1} \\ \vdots & \ddots & \vdots \\ \psi_{1N} & \cdots & \psi_{LN} \end{pmatrix}, \quad (1)$$

$$\Phi = \begin{pmatrix} \phi_{11} & \cdots & \phi_{L1} \\ \vdots & \ddots & \vdots \\ \phi_{1M} & \cdots & \phi_{LM} \end{pmatrix}, \quad (2)$$

where $\psi_{ln}$ is the average WiFi RSS reading at WiFi AP $n$ from a dual band UE located at LP $l$. In order to compensate the changes in the UE transmitted power, these WiFi RSS measurements are normalized by the UE transmit power in the 5 GHz band. $L$ is the total number of LPs, and $N$ is the total number of WiFi APs. $\phi_{lm}$ is equal to the best sector ID number corresponding to the maximum power received by a dual band UE located at LP $l$ from mm-w AP $m$, where $M$ is the total number of mm-w APs. $\phi_{lm}$ can be calculated as:

$$\phi_{lm} = d_m^* = \arg\max_{d_m}\left(P_{lm}(d_m)\right), \quad 1 \leq d_m \leq D_m, \quad (3)$$

where $d_m$ indicates the sector ID number of mm-w AP $m$, $D_m$ is the total number of sectors IDs of mm-w AP $m$, $d_m^*$ is the best sector ID number corresponding to the maximum received power at LP $l$ from mm-w AP $m$, and $P_{lm}(d_m)$ indicates the power received at LP $l$ from mm-w AP $m$ using sector ID $d_m$. We assume that all LPs are under the coverage of the $N$ WiFi APs.

A *null* value in the $\Phi$ matrix, i.e., $\phi_{lm} = null$, means that mm-w AP $m$ cannot cover LP $l$. A *null* value is decided for $\phi_{lm}$ if no signal is detected at LP $l$ after mm-w AP $m$ scans all its antenna sectors, or the maximum received power is less than the power required for the communication using the lowest Modulation Coding Scheme (MSC) index. *null* best sector IDs can be used to solve WiFi and mm-w association / re-association problem beforehand using the statistical learning information, which is left as our future work.

Figure 4 shows an example of a mm-w AP best sector ID radio map using uniformly distributed LPs in a room area of 200 m². In this example, the mm-w AP is located at X= 14 m, Y = 3.5 m and Z= 3 m. The color bar indicates the best sector ID number. Each square in Fig. 4 indicates a different LP, and each LP is covered by a certain mm-w best sector ID, and it has a certain WiFi RSS readings.

### 2) Grouping and Clustering the WiFi Fingerprints

In the proposed BF scheme, offline WiFi fingerprints are mainly used to localize mm-w best sector IDs using LPs. Because many of the LPs may be covered by the same best sector ID $d_m^*$ while their WiFi RSS readings are different, as it is shown in Fig. 4, the WiFi RSS readings from the LPs covered by the same best sector ID $d_m^*$ are grouped together.

$$\xi_k^{d_m^*} = \psi_l|_{\forall l \in \{\phi_{lm}=d_m^*\}}, \quad 1 \leq k \leq K_{d_m^*}, \quad (4)$$

where $\psi_l = [\psi_{l1}\ \psi_{l2}\ \ldots\ \psi_{lN}]^T$ is the WiFi RSS readings vector from LP $l$. $\xi_k^{d_m^*}$ is the $k$-th WiFi RSS vector in group $d_m^*$. $K_{d_m^*}$ is the total number of WiFi RSS vectors in group $d_m^*$, which is equal to the total number of LPs that can be covered by $d_m^*$. For example, in Fig. 4, all WiFi RSS vectors from the LPs covered by mm-w best sector ID 15 (yellow squares) are grouped together.

To reduce the area of interest and the computational complexity of WiFi RSS matching in the online BF phase, clustering is conducted on each group of WiFi RSS vectors. Accordingly, the best WiFi RSS exemplars that can effectively represent each best sector ID $d_m^*$ are obtained. Because mm-w transmissions are highly affected by shadowing and reflections, the radio map of mm-w best sector IDs is irregular and highly overlapped, as it is shown in Fig. 4. Therefore, an efficient clustering algorithm should consider this overlapping nature during forming the clusters. The widely used K-means clustering algorithm [10] cannot be directly applied to the proposed BF mechanism. This is because in the traditional K-means clustering, a cluster exemplar is the centroid of the nearest K vectors, which may not necessarily be a member of the data set. Thus, by directly applying the K-means algorithm for clustering the WiFi RSS fingerprints in a best sector ID group, some of the calculated WiFi RSS exemplars may belong to other best sector IDs due to the aforementioned overlapping nature. Instead, in this paper, we use the affinity propagation clustering algorithm [9] as an appropriate clustering algorithm for the proposed BF protocol. Affinity propagation is a fast clustering algorithm that simultaneously considers all WiFi RSS vectors in the same group as potential exemplars for clusters formation by assigning the same preference value for all of them. This property is highly favored by the proposed BF due to the overlapping nature of the best sector ID radio map. The similarity indicator $s(i,k)$ indicates how well the WiFi RSS vector $k$, $\xi_k^{d_m^*}$, is suited to be the exemplar for WiFi RSS vector $i$, $\xi_i^{d_m^*}$, where $\xi_k^{d_m^*}$ and $\xi_i^{d_m^*}$ belong to the same group $d_m^*$. $s(i,k)$ can be calculated as [9]:

$$s(i,k) = -\left\|\xi_i^{d_m^*} - \xi_k^{d_m^*}\right\|^2, \quad \forall i,k \in \{1,2,\ldots,K_{d_m^*}\}, k \neq i. \quad (5)$$

The self-similarity value $s(k,k)$ indicates the preference value of $\xi_k^{d_m^*}$ to be a cluster exemplar. Because all WiFi RSS vectors have equal potentials to be cluster exemplars, $s(k,k)$ is set as [9]:

$$s(k,k) = \chi \cdot \text{median}\{s(i,k), \forall i,k \in \{1,2,\ldots,K_{d_m^*}\}, k \neq i\}, \quad (6)$$

where $\chi$ is a design parameter, which is experimentally determined to control the number of generated clusters.

The core operation of the affinity propagation algorithm is the exchange of two-kind of messages in a recursive manner. These messages are the responsibility message $r(i,k)$ and the availability message $a(i,k)$. The responsibility message $r(i,k)$ is sent from $\xi_i^{d_m^*}$ to the candidate exemplar $\xi_k^{d_m^*}$ to know how well-suited $\xi_k^{d_m^*}$ is to serve as the exemplar for $\xi_i^{d_m^*}$. The availability message $a(i,k)$ is sent from the candidate exemplar $\xi_k^{d_m^*}$ to $\xi_i^{d_m^*}$ to reflect the suitability that $\xi_k^{d_m^*}$ becomes the exemplar of $\xi_i^{d_m^*}$. $r(i,k)$ and $a(i,k)$ can be defined as follows [9]:

$$r(i,k) = s(i,k) - \max_{\grave{k} \text{ s.t. } \grave{k} \neq k}\{a(i,\grave{k}) + s(i,\grave{k})\}, \quad (7)$$

$$a(i,k) = \min\left\{0, r(k,k) + \sum_{\grave{i} \text{ s.t. } \grave{i} \neq i} \max\{0, r(\grave{i},k)\}\right\}, \quad (8)$$

$$a(k,k) = \sum_{\grave{i} \text{ s.t. } \grave{i} \neq k} \max\{0, r(\grave{i},k)\}. \quad (9)$$

At any iteration during the affinity propagation process, availabilities and responsibilities can be combined to identify the clusters exemplars and their associated members. At the end of the affinity propagation clustering process, we obtain the set of WiFi RSS exemplars $\mathfrak{I}_j^{d_m^*}, j \in \{1,2,\ldots,C_{d_m^*}\}$, for best sector ID $d_m^*$, where $C_{d_m^*}$ is the total number of WiFi RSS exemplars (clusters) for $d_m^*$. For example, in Fig. 4, four WiFi RSS exemplars are calculated using the affinity propagation algorithm to effectively localize and represent the best sector ID 15, and Fig. 4 shows the LPs corresponding to these exemplars.

The offline statistical learning phase will not be repeated unless the transmit power and location of the APs are changed, or the internal structure of the target environment is changed.

### B. The Online BF Phase

The proposed real-time BF protocol mainly consists of 3- steps, the online WiFi RSS measurements, the estimation of best beams and the BC subphase. The actual BF protocol takes place in this real-time phase.

*1) Collecting the Current WiFi RSS Readings*

During this step, the online WiFi RSS vector $\psi_r$ is measured by the WiFi APs and collected by the controller from the dual band UE located at an arbitrary position $r$. $\psi_r$ is defined as:

$$\psi_r = [\psi_{r1}\,\psi_{r2}\,\ldots\,\psi_{rN}]^T. \quad (10)$$

*2) The Best Beams Estimation*

The second step in the online BF protocol, after collecting the online WiFi RSS readings, is to estimate a group of best beams for each mm-w AP-UE link. This can be done by calculating the smallest Euclidian distance between $\psi_r$ and the WiFi RSS exemplars of each best sector ID $d_m^*$, i.e., the Euclidian distance corresponding to the nearest WiFi RSS exemplar to $\psi_r$ from each best sector ID $d_m^*$. Therefore, a vector of smallest Euclidian distances is obtained with a length up to the total number of best sector IDs. The controller sorts the obtained vector of smallest Euclidian distances in an ascending order, and it selects a group of best sector IDs (best beams) $d_m^*(1:X)$ corresponding to the nearer WiFi RSS exemplars to $\psi_r$ from different best sector IDs. These best beams $d_m^*(1:X)$ are used for the BC subphase.

$$d_m^*(1:X) = \text{sort}_{d_m^*}\left(\arg\min_{1 \leq j \leq C_{d_m^*}}\left\|\psi_r - \mathfrak{I}_j^{d_m^*}\right\|^2\right)\bigg|_{1:X}. \quad (11)$$

*3) The Beamforming Combining (BC) Subphase*

The BC subphase is mainly used by the proposed BF protocol to overcome the imperfect quasi-omni antenna pattern, to alleviate the real-time beam blocking, e.g., human shadowing, and to overcome the inaccuracy in the WiFi RSS measurements. After selecting the best $X$ beams for mm-w AP $m$, $d_m^*(1:X)$, to communicate with the UE at its current position $r$, the controller sends the IDs of the selected beams to the mm-w AP. If a mm-w AP wants to communicate with the UE, it should perform a BC subphase with the UE in the form of beam training using the estimated beams before data transmission. After the BC training, the beam pattern corresponding to the highest link quality will be selected as the best beam for a mm-w AP-UE link.

The BC training can be formulated as:

$$d_m^*(x^*) = \arg\max_x(P_{rm}(d_m^*(x))), 1 \leq x \leq X, \quad (12)$$

where $d_m^*(x^*)$ is the best beam ID number for mm-w AP $m$–UE link, and $P_{rm}(d_m^*(x))$ is the power received by the UE at its current position $r$ from mm-w AP $m$ using sector ID $d_m^*(x)$. Instead of only estimating the best beam for a mm-w AP-UE link, it is better to reserve the higher link quality beams, so that fast beam switching can be accomplished in the case of instantaneous link blockage.

### C. The Frame Format of the Proposed Dual Band BF Protocol

Figure 5 shows the frame format of the proposed BF protocol, for one mm-w AP-UE scenario. In this frame format a dual band (5 GHz / 60 GHz) MAC protocol is used. For the sake of simple explanation, the minor details of the MAC protocol, i.e., inter-frame durations, random backoff… etc., are not shown in Fig. 5. A probe request (Prob. Req.) management packet is frequently announced by the UE using its 5 GHz interface for WiFi RSS measurements. $T_{RSS}$ is the time required for sending the Prob. Req. packet, and $T_{PT}$ is the processing time required by the controller to estimate the group of mm-w best beams based on the current WiFi RSS readings. $T_{PT}$ includes the time required for collecting the current WiFi RSS readings from the WiFi APs, the time required for Euclidian distance calculations and the time required for sending the estimated best beams IDs to the mm-w AP. Using these estimated beams, a BC subphase using Beam Refinement Packets (BRPs) is performed by the mm-w AP to find out the best beam for the mm-w AP-UE link. At the end of the BC subphase, a Feedback (FB) packet is sent from the UE to the mm-w AP, using its 60 GHz interface, to report the ID number of the highest link quality TX beam.

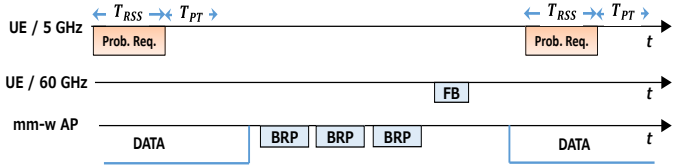

Fig. 5. The frame format of the proposed dual band mm-w BF protocol.

If the UE has many receive antenna sectors, an MID subphase should be performed before the BC subphase to find out the best TX / RX antenna beams.

In order to reduce the latency of the proposed scheme, thanks to the orthogonality between 5 GHz and 60 GHz bands, we propose that the UE announces the 5 GHz Prob. Req. packet and the controller estimates the best beams during mm-w data transmission. Thus, during the current time slot of mm-w data transmission, the UE sends the 5 GHz Prob. Req. packet and the controller estimates the best beams for the BC subphase of the next time slot of mm-w data transmission. Therefore, the latency of the proposed scheme is highly reduced. If the time slot period for mm-w data transmission is properly adjusted, the only setup time required by the proposed BF protocol will be the BC time. If an MID subphase is used, the total setup time becomes the MID and BC times. Accordingly, The BF setup time required by the proposed protocol is less than the time required by the IEEE 802.11ad protocol by the SLS time, which consumes more than 70 % of the total BF time.

## IV. SIMULATION ANALYSIS

In this section, the efficiency of the proposed BF protocol compared to the exhaustive search IEEE 802.11ad BF protocol is verified via computer simulations. In the simulation analysis, a performance metric of the power received from the best beam estimated by the proposed BF normalized to the power received from the best beam estimated by the IEEE 802.11ad BF is used.

### A. Simulation Area and Simulation Parameters

Figure 6 shows the ray tracing simulation area of an indoor environment. For the purpose of generalization, we consider separate deployments of the WiFi APs from the mm-w APs. Room materials are from concrete except the desks are made of wood. Other simulation parameters are given in Table I.

The steering antenna model, which is defined in the IEEE 802.11ad [7], is used as the transmit antenna directivity for the mm-w AP, in which the 3D beam gain in dB can be defined as follows:

$$G(\varphi,\theta)[dB] = G_0[dB] - \min[-(G_H(\varphi) + G_V(\theta)), A_m], \quad (13)$$

$$A_m[dB] = 12 + G_0[dB], \quad (14)$$

$$G_0[dB] = 20\log_{10}\left(\frac{1.6162}{\sin\left(\frac{\theta_{-3dB}}{2}\right)}\right), \quad (15)$$

where $\varphi$, $\theta$ are the azimuth and elevation angles, and $G_H(\varphi)$, $G_V(\theta)$ are the beam gains in the horizontal and vertical directions, which can be defined as:

$$G_H(\varphi) = -\min\left[12\left(\frac{\varphi - \varphi_{beam}}{\varphi_{-3dB}}\right)^2, A_m\right], \quad (16)$$

$$G_V(\theta) = -\min\left[12\left(\frac{\theta - \theta_{tilt}}{\theta_{-3dB}}\right)^2, A_m\right], \quad (17)$$

where $\varphi_{-3dB}$ and $\theta_{-3dB}$ are the half power beamwidths in the horizontal and vertical directions. $\varphi_{beam}$ and $\theta_{tilt}$ are the angles corresponding to the beam center.

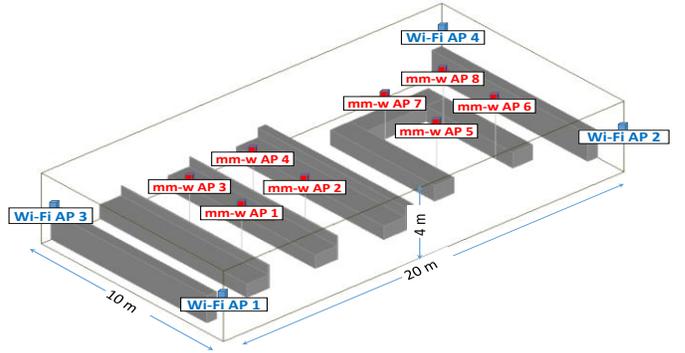

Fig. 6. The ray tracing simulation area.

TABLE .1 THE SIMULATION PARAMETERS

| Parameter | Value |
| --- | --- |
| Number of WiFi APs | 4 |
| Number of mm-w APs | 8 |
| Tx power of WiFi APs / mm-w APs | 20dBm /10dBm |
| $\theta_{-3dB}$, $\varphi_{-3dB}$ for mm-w AP beam | 20° |
| Antenna model for the 5GHz band | Omni |
| Number of antenna sectors for mm-w AP / UE | 92 / 1 |
| Antenna model of the UE @ 60 G Hz | Quasi-omni |
| Probability of 60 GHz Line of Sight (LOS) path blocking due human shadowing [7] | Uniform distribution |
| χ (the clusters generation number) | 0.3 |

In this paper, we consider 3D beamforming in order to fully cover the indoor area, i.e., each beam has different values of $\varphi_{beam}$ and $\theta_{tilt}$. Thus, the total channel gain from mm-w AP $m$ to UE $k$ becomes:

$$g_{mk}(\tau) = \int_0^{2\pi}\int_0^{\pi} \sqrt{G_m(\varphi,\theta)} h_{mk}(\varphi,\theta,\tau) \sin\theta d\theta d\varphi, \quad (18)$$

where $h_{mk}(\varphi,\theta,\tau)$ is the complex channel response between mm-w AP $m$ and UE $k$ without beamforming gain.

### B. Simulation Results

Figure 7 shows the average Received Power Ratio (RPR) in dB between the power received from the best beam estimated by the proposed BF protocol and the power received from the best beam estimated by the exhaustive search IEEE 802.11ad BF using different number of LPs. The average is taken over the used number of mm-w APs (8 mm-w APs). In this simulation, we use 4-WiFi APs, as it is shown in Fig. 6. Also, 5 best beams are estimated for the BC subphase. For the purpose of comparison, we give the performance of only using the nearest neighbor (N. N.) best beam. In this scheme, based on the current WiFi RSS readings vector, the controller calculates its N. N. offline WiFi RSS vector from the entire WiFi RSS fingerprints database. Then, it selects mm-w best sector ID corresponding to this N. N. WiFi RSS vector as the best beam for the mm-w AP-UE link. Thus, grouping and clustering are not performed on the offline WiFi RSS fingerprints. Also, no real-time beam training is used.

From Fig. 7, as the number of LPs is increased, the performance of the proposed BF protocol is enhanced. This is because, as the number of LPs is increased, high resolution micro wave and mm wave radio maps can be constructed for the indoor environment. Accordingly, the proposed BF can accurately estimate the best beam for a mm-w AP-UE link. Also, the performance of estimating a group of best beams for real-time beam training is better than only selecting the N. N. offline best beam.

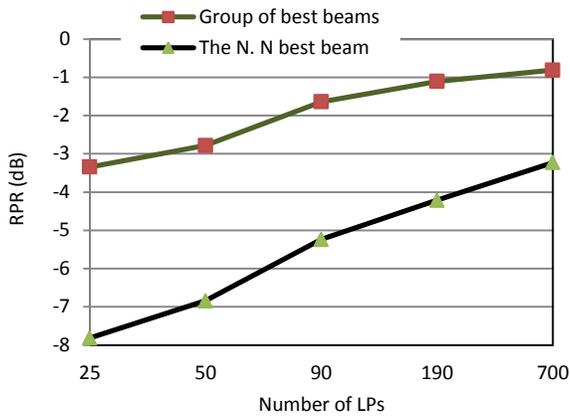

Fig. 7. The average RPR using different number of LPs.

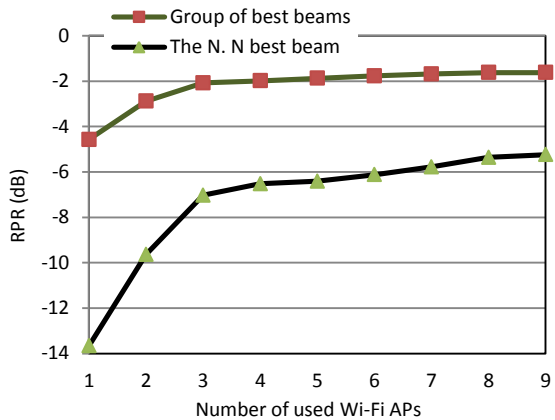

Fig. 8. The average RPR using different number of WiFi APs.

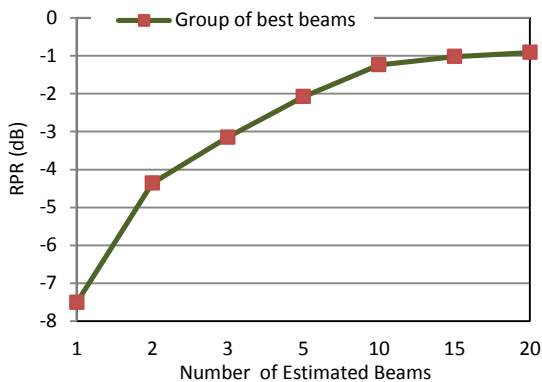

Fig. 9. The average RPR using different number of estimated beams.

This is because the real-time beam training overcomes the online link blockage, and it compensates the fluctuations in the WiFi RSS measurements. These problems cannot be handled by only estimating the N. N. offline best beam; simply because the estimated offline beam may be blocked in the real-time scenario. From the trade-off between complexity and performance, 90 LPs are selected as a sufficient number of LPs.

Figure 8 shows the average RPR performance of the proposed BF protocol using different number of WiFi APs. The WiFi APs are distributed one by one in room corners and centers. In this simulation, we use 90 LPs and 5 best beams. As the number of WiFi APs is increased, the RPR performance is enhanced. This is because the accuracy of the WiFi RSS fingerprints is increased. From the trade-off between deployment cost and performance 3 WiFi APs are selected as a sufficient number of WiFi APs.

Figure 9 shows the average RPR performance of the proposed BF protocol when increasing the number of estimated best beams for real-time beam training. In this simulation, 90 LPs and 3-WiFi APs are used. As it is clearly shown, as we increase the number of estimated best beams, the RPR performance is enhanced. By only using 10 training beams, only 1 dB difference in maximum received power from the exhaustive search BF is obtained using the proposed BF.

## V. CONCLUSION

In this paper, we focused on the problem of developing a fast mm-w BF protocol with a comparable performance to the conventional exhaustive search BF as an enabler for applying the 60 GHz mm-w technology in 5G networks. In this regard, we proposed a novel mm-w BF mechanism based on a statistical learning approach. In this mechanism, the best beam for a mm-w AP-UE link is estimated using WiFi fingerprints. The micro-wave and mm-wave radio maps are pre-constructed for target environments in an offline phase. Consequently, the best beam for a mm-w AP-UE link can be immediately estimated by just comparing the current UE WiFi readings with the pre-stored WiFi fingerprints. Using 92 antenna sectors, 10 estimated best beams, 3 WiFi APs and 90 LPs, the proposed BF protocol succeeded to estimate the best beam for a mm-w AP-UE link with a decrease in BF setup time of more than 70 % compared to the exhaustive search BF. This high decrease in BF setup time comes at the expense of only 1 dB decrease in maximum received power. The proposed BF protocol can be used as an enabler for coordinated mm-w concurrent transmissions and its extension for 5G cellular networks. For future work, dynamic learning will be investigated in conjunction with the proposed BF protocol for further enhancements of its complexity and performance.

ACKNOWLEDGMENT

This work is partly supported by "Research and development project for expansion of radio spectrum resources" of MIC, Japan.